\documentclass{ws-ijmpd}

\begin{document}

\markboth{Wayne de Paula} {Scalar Spectrum from a Dynamical Gauge/Gravity model.}

%
\catchline{}{}{}{}{}
%

    \title{Scalar Spectrum from a Dynamical Gravity/Gauge model.}

    \author{W de Paula}

    \address{Departamento de F\'\i sica, Instituto Tecnol\'ogico de Aeron\'autica,
    12228-900 S\~ao Jos\'e dos Campos, SP, Brazil.\\wayne@ita.br}

    \author{T Frederico}

    \address{Departamento de F\'\i sica, Instituto Tecnol\'ogico de Aeron\'autica,
    12228-900 S\~ao Jos\'e dos Campos, SP, Brazil.\\
    tobias@ita.br}

    \maketitle


\begin{abstract}
We show that a Dynamical AdS/QCD model is able to reproduce the
linear Regge trajectories for the light-flavor sector of mesons
with high spin and also for the scalar and pseudoscalar ones. In
addition the model has confinement by the Wilson loop criteria and
a mass gap. We also calculate the decay amplitude of scalars into
two pion in good agreement to the available experimental data.
\end{abstract}


\section{Introduction}

Over the years experiments confirm that strong interaction is
successfully described by Quantum Chromodynamics (QCD). For very
high energies one can calculate physical amplitudes analytically
using the QCD Lagrangian due to asymptotic freedom. On the other
hand we have a lack of analytical tools to analyze the low energy
sector. Important properties of the infrared physics of the strong
interaction such as confinement, mass gap and linear Regge
trajectories remains unexplained by QCD.

In $1974$ 't Hooft proposed a duality between the large N (number
of colors) limit of QCD and string theory \cite{tHooft}. This
represented the first dual representation of a gauge theory by a
string model. In $1998$ Maldacena\cite{Maldaconj} proposed a
mapping between operators in Conformal Field Theory (CFT) and
fields of a $\it{N}=4$ Type IIB string field theory in a
ten-dimensional space-time $AdS_{5}\times S_{5}$. The most
interesting fact of this duality is that the strong-coupling
regime of large-$N_{c}$ gauge theories can be approximated (in
low-curvature regions) by weakly coupled and hence analytically
treatable classical gravities. The drawback is that CFT is not
QCD. Consequently, the $\it{N}=4$ Type IIB string field in
$AdS_{5}\times S_{5}$ does not have many important properties of
strong interactions as confinement and a mass gap.

A direct way for searching for a QCD dual is introducing D-branes in the theory. They are responsible for breaking in part supersymmetry and for
the introduction of flavor. For example, the addition of $N_{f}$ D7 probe branes ($D3-D7$ model\cite{D3D7}) can be interpreted as the introduction
of flavor in the AdS/CFT. In the supergravity side it is a four-dimensional $\it{N} = 2$ supersymmetric large-N gauge theory. Although a Type II B
$\it{N} = 2$ has a running coupling constant, it does not has confinement. There is a vast literature addressing these topics  and for a review
see \cite{Erdmenger}. In all those models (top-down) we obtain a one-dimensional differential equation in holographic coordinate to calculate the
mass spectra and they do not lead to a Regge spectrum for meson excitations (see e.g. \cite{dePaulaBianchi}). This fact suggested an other way of searching for the corresponding
QCD dual. We propose an effective 5d action that can reproduce basic properties of strong interaction and we explore the phenomenological aspects
of this model in a bottom-up approach.

The first model with this idea was proposed by Polchinski and Strassler\cite{Polchinski}. This model (hard-wall) is a slice of AdS with an IR
boundary condition that introduces the QCD scale. It implements the counting rules which govern the scaling behavior of hard QCD scattering
amplitudes by the conformal invariance of AdS$_{5}$ in the UV limit. In spite of reproducing a large amount of hadron phenomenology\cite{hwph} it
does not have linear Regge trajectories. A soft-wall model\cite{Karch} was created to correct this problem, where the AdS$_{5}$ geometry is kept
intact while an additional dilaton background field is introduced. This dilaton soft-wall model indeed generates linear Regge trajectories
$m_{n,S}^{2}\sim n+S$ for light-flavor mesons of spin $S$ and radial excitation level $n$. (Regge behavior can alternatively be encoded via IR
deformations of the AdS$_{5}$ metric \cite{Kruczenski,FBT}.) However, the resulting vacuum expectation value (vev) of the Wilson loop in the
dilaton soft wall model does not exhibit the area-law behavior in contrast to a linearly confining static quark-antiquark potential. It happens
because the model uses an AdS metric which is not of a confining type by the Wilson loop analysis \cite{Malda2,Rey01}. In addition the soft-wall
model background is not a solution of a dual gravity. Csaki and Reece \cite{Csaki} analyzed the solutions of a 5d dilaton-gravity Einstein
equations (see also\cite{Kiritsis}) using the superpotential formalism. They concluded that it would not be possible to solve those equations and
obtain a linear confining background without introducing new ingredients. They suggested to analyze a tachyon-dilaton-graviton model, and this
idea was successfully implemented in \cite{Batell}.

We took an alternative route and we show\cite{dePaulaPRD09} that we can obtain a linear confining background as solution of the dilaton-gravity
coupled equations.  Within our proposal of a self-consistent dilaton-gravity model, the mass spectrum of the high spin mesons stays close to a
linear Regge trajectory for the lower excitations, where experimental data exists, while an exact linear behavior is approached for high spin and
mass excitations.(Using similar approach in \cite{Pirner} is proposed a different AdS deformation in order to reproduce the QCD running coupling.)
\subsection{Hadronic Resonances in Dynamical AdS/QCD model}
The action for five-dimensional gravity coupled to a dilaton field
is:
\begin{eqnarray}
S = \frac{1}{2k^{2}}\int d^{5}x \sqrt{g} \left( -\emph{R} - V(\Phi)+\frac{1}{2}g^{MN}\partial_{M}\Phi\partial_{N}\Phi\right), \label{actiongd}
\end{eqnarray}
\noindent where $k$ is the Newton constant in $5$ dimensions and
$V(\Phi)$ is the scalar field potential. We will be restricted to
the metric family: $g_{MN}=e^{-2A(z)}\eta_{MN}$, where $\eta_{MN}$
is the Minkowski one. Minimizing the action, we obtain a coupled
set of Einstein equations for which the solutions satisfy the
following relations:
\begin{equation}
\Phi'= \sqrt{3 A'^{2} + 3 A''}~,~V(\Phi) = \frac{3 e^{2A}}{2} \left(A''-3 A'^{2}\right). \label{constrain}
\end{equation}
The $5$d action for a gauge field $\phi_{M_{1}\dots M_{S}}$ of spin $S$ in the background is given by\cite{Karch}
\begin{equation}
I = \frac{1}{2} \int d^{5}x \sqrt{g} e^{-\Phi}\left(\nabla_{N} \phi_{M_{1}\dots M_{S}} \nabla^{N} \phi^{M_{1}\dots M_{S}} \right).
\end{equation}
As in \cite{Karch} and \cite{KatzLewandowski}, we utilize the
axial gauge. To this end, we introduce new spin fields
$\widetilde{\phi}_{\dots}= e^{2(S-1)A}\phi_{\dots}$. We also make
the substitution $\tilde{\phi_{n}}= e^{B/2}\psi_{n}$ and obtain a
Sturm-Liouville equation
\begin{equation}
\left(-\partial_{z}^{2}+ {\mathcal V}_{eff}(z)\right)\psi_{n} = m_{n}^{2}\psi_{n},
\end{equation}
\noindent where $B = A (2S-1) + \Phi$ and ${\mathcal
V}_{eff}(z)=\frac{B'^{2}(z)}{4} -\frac{B''(z)}{2}$. Hence, for
each metric $A$ and dilaton field $\Phi$ consistent with the
solutions of the Einstein equations, we obtain a mass spectrum
$m_n^2$. Due to the gauge/gravity duality this mass spectrum
corresponds to the mesonic resonances in the $4$d space-time.

Now we will focus on scalar mesons (also analyzed in
\cite{Schmidt_Scalar,Colangelo_Scalar}). The
action\cite{dePaulaArXiv09}
\begin{eqnarray}
I = \frac{1}{2} \int d^{4}x dz\sqrt{\left\vert g\right\vert } \left(g^{\mu\nu}\partial_{\mu}
\varphi(x,z)\partial_{\nu}\varphi(x,z)-\frac{M_{5}^{2}}{\Lambda_{QCD}^{2}}\varphi^{2} \right),
\end{eqnarray}
describes a scalar mode propagating in the dilaton-gravity
background. Factorizing the holographic coordinate dependence as
$\varphi(x,z)=e^{iP_{\mu}x^{\mu}}\varphi(z)$ with
$P_{\mu}P^{\mu}=m^{2}$, and redefining the string amplitude as
$\psi _{n}(z)=\varphi_{n}(z) \times e^{-(3A+\Phi)/2}$, we have a
Sturm-Liouville equation
\begin{equation}
\left[ -\partial _{z}^{2}+\mathcal{V}(z)\right] \psi _{n}=m_{n}^{2}\psi _{n},  \label{sleq}
\end{equation}%
where the string-mode potential is $\mathcal{V}(z)=\frac{B^{\prime
}{}^{2}(z)}{4}-\frac{B^{\prime \prime
}(z)}{2}+\frac{M_{5}^{2}}{\Lambda_{QCD}^{2}}e^{-2A(z)}$ with
$B=3A+\Phi$. (Note that $B=(2S-1)A+\Phi$ for the spin nonzero
states \cite{Karch}.) The AdS/CFT correspondence states that the
wave function should behave as $z^{\tau}$, where $\tau = \Delta -
\sigma$ (conformal dimension minus spin) is the twist dimension
for the corresponding interpolating operator that creates a given
quark-gluon configuration \cite{Polchinski}. The five-dimensional
mass chosen as \cite{Witten98} $M_{5}^{2}=\tau(\tau-4),$ fixes the
UV limit of the dual string amplitude with the twist dimension.
\section{Phenomenological Results}
Our aim was to construct a metric ansatz that is AdS in the UV and
allows for confinement, mass gap and Regge trajectories by
tailoring its IR behavior. With these constraints we found the
metric:
\begin{equation}
A(z)= Log(\xi z \Lambda_{QCD}) + \frac{(\xi z\Lambda _{QCD})^{2}}{1+e^{(1-\xi z\Lambda _{QCD})}} \label{cnew}
\end{equation}
For scalars $\xi = 0.58$. To distinguish the pion states in our
model, the fifth dimensional mass was rescaled according to
$M_{5}^{2} \rightarrow M_{5}^{2}+\lambda z^{2}$ (see \cite{FBT}).
The model is constrained by the pion mass, the slope of the Regge
trajectory and the twist 2 from the operator $\bar q \gamma^5 q$.
The results for the Regge trajectories for $f_0$ and pion are
shown in figure 1. The pion modes are calculated with $\xi=$0.88
and $\lambda=-2.19$GeV$^2$. For high spin mesons, see figure 2, we
have an equation to obtain the scale factor $\xi=S^{-0.3329}$ in
order to keep the slope of the Regge trajectories fixed. (Note
that in our previous work\cite{dePaulaPRD09} we adopted a slightly
different ansatz.)
\begin{figure}[tbh] \centerline{\epsfig{figure=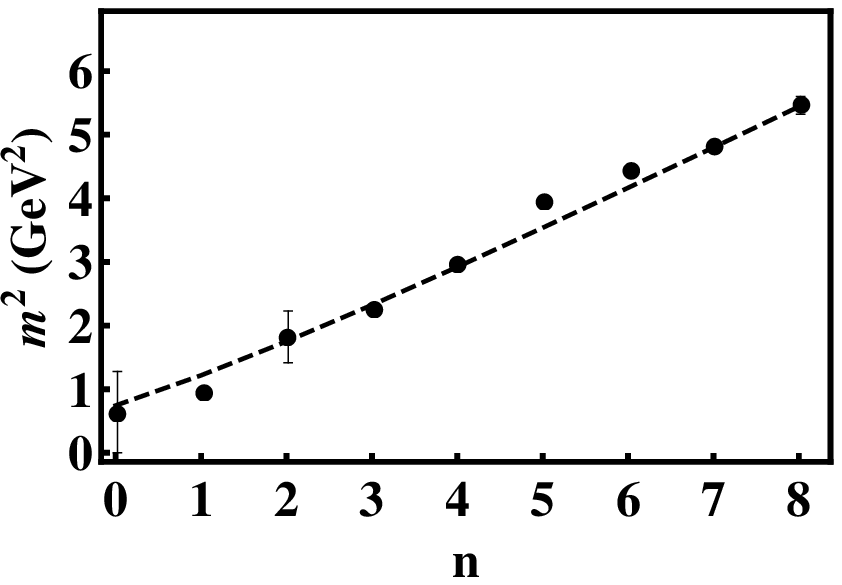,width=4.3cm,height=3.4cm} \epsfig{figure=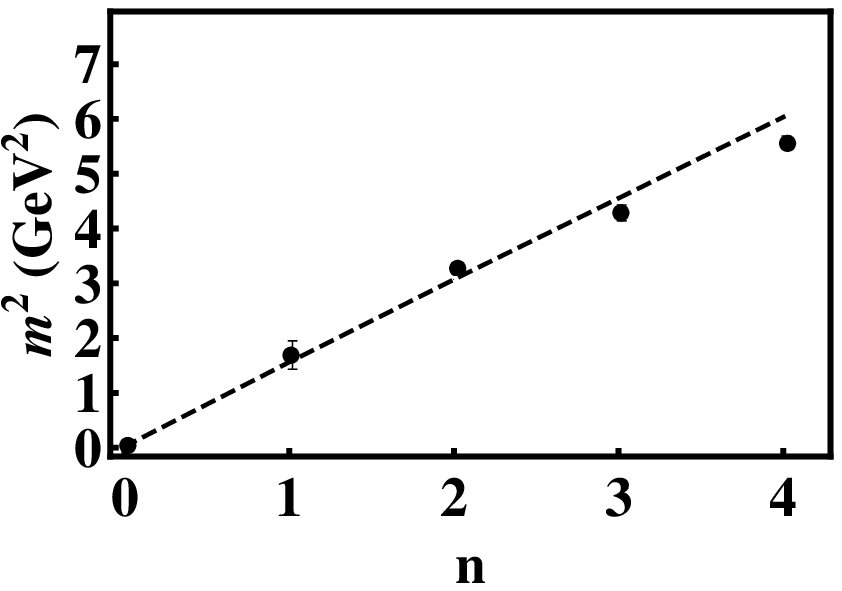,width=4.3cm,height=3.4cm}}
\caption{Regge trajectory for $f_0$ (left panel) and pion (right
panel) from  the Dynamical AdS/QCD model with $\Lambda _{QCD}=0.3$
GeV. Experimental data from PDG. } \label{Fig1}
\end{figure}
\begin{figure}[tbh]
\vspace{-0.5cm}
\centerline{\epsfig{figure=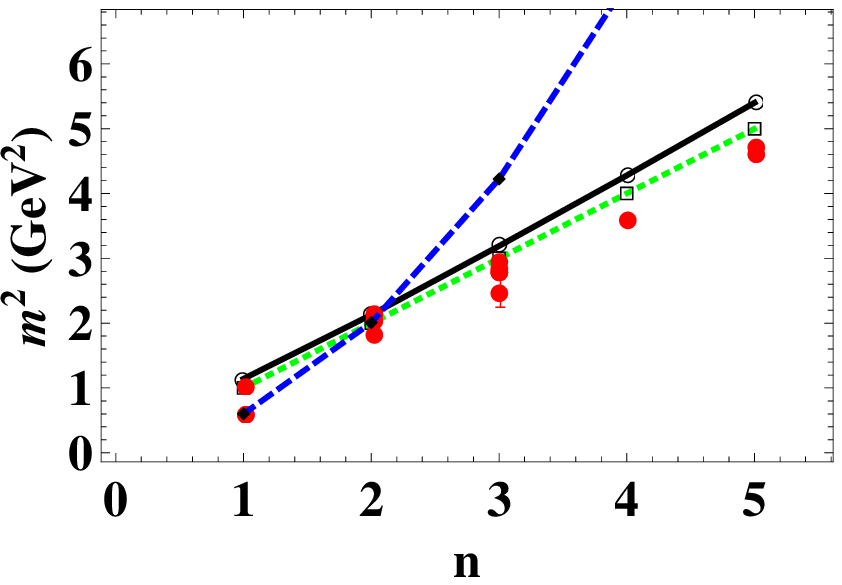,width=4.3cm,height=3.4cm}
\epsfig{figure=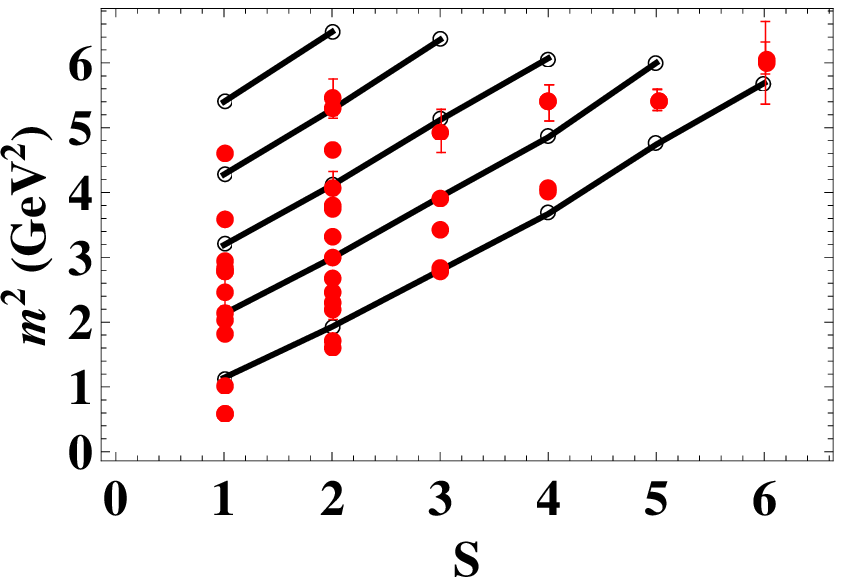,width=4.3cm,height=3.4cm}}
\caption{Radial excitations of the rho meson in the hard-wall
(dashed line), soft-wall \protect\cite{Karch} (dotted line) and
our dynamical soft-wall (solid line, for $\Lambda _{QCD}=0.3$ GeV)
backgrounds (left panel). Dynamical AdS/QCD spectrum for spin
excitations (right panel). Experimental data from PDG.}
\label{Fig2}
\end{figure}

\vspace{-0.8cm}
\section{Decay Amplitudes}
The $f_0$'s partial decay width into $\pi\pi$ are calculated from
the overlap integral ($h_n$) of the normalized string amplitudes
(Sturm-Liouville form) in the holographic coordinate  dual to the
scalars $(\psi_n)$ and pion $(\psi_\pi)$ states,
\begin{equation}
h_{n} =k\int^\infty_0 dz ~\psi_\pi^2(z)\psi_n(z) \ , \text{~~~
with~~~} \int_{0}^{\infty}dz\psi_{m}(z)\psi_{n}(z)=\delta_{mn} \ .
\label{dwover}
\end{equation}
The constant $k$ has dimension $\sqrt{mass}$ fitted to the
experimental value of the $f_0(1500)\to \pi\pi$ partial decay
width. The Sturm-Liouville amplitudes of the scalar (pseudoscalar)
modes are normalized just as a bound state wave function in
quantum mechanics \cite{RadyushkinPRD2007,BrodskyPRD2008025},
which also corresponds to a normalization of the string amplitude.
The overlap integral for the decay amplitude, $h_n$, is the dual
representation of the transition amplitude $S\to PP$ and therefore
the decay width is given by $\Gamma_{\pi \pi}^{n} =
\frac{1}{8\pi}|h_{n}|^{2}\frac{p_\pi}{m_n^2}\ ,$ where $p_\pi$ is
the pion momentum in the meson rest frame.

The known two-pion partial decay width for the $f_0$'s given in the particle listing of PDG\cite{pdg}, are calculated with Eq. (\ref{dwover}) and
shown in Table I. The width of $f_0(1500)$ is used as normalization.  In particular for $f_0(600)$ the model gives a width of about 500 MeV, while
its mass is 860 MeV. The range of experimental values quoted in PDG for the sigma mass and width are quite large as depicted in Table I. The
analysis of the E791 experiment gives $m_\sigma=$ 478$^{+24}_{-23}\pm 17$ MeV and $\Gamma_\sigma=324^{+42}_{-40}\pm 21$ MeV \cite{Aitalasigma}.
The width seems consistent with our model while the experimental mass appears somewhat smaller. The CLEO collaboration \cite{CLEO2002} quotes
$m_\sigma=$ 513$\pm 32$ MeV and $\Gamma_\sigma=335\pm 67$ MeV, and a recent analysis of the sigma pole in the $\pi\pi$ scattering amplitude from
ref.\cite{Caprini06} gives $m_\sigma=$ 441$^{+16}_{-8}$ MeV and $\Gamma_\sigma=544^{+18}_{-25}$ MeV. Other analysis of the $\sigma$-pole in the
$\pi\pi \to \pi\pi$ scattering amplitude present in the decay of heavy mesons indicates a mass around 500 MeV \cite{bugg06}. \vspace{-0.3cm}
\begin{table}[tph]
\tbl{Two-pion decay width and masses for the $f_{0}$
family.Experimental values from PDG. $^\dagger$Mixing angle of
$20^o$. $^*$Fitted value.} {\begin{tabular}{@{}ccccc@{}} \toprule
\hline
Meson         &$M_{exp}$(GeV) & $M_{th}$(GeV) & $\Gamma^{exp}_{\pi\pi}$(MeV) & $\Gamma^{th}_{\pi\pi}$(MeV) \\
\hline
$f_{0}(600) $ & 0.4 - 1.2        & 0.86   & 600 - 1000    & 535 \\
$f_{0}(980) $ & 0.98$\pm$ 0.01   & 1.10   & $\sim$15-80   & 42$^\dagger$  \\
$f_{0}(1370)$ & 1.2 - 1.5        & 1.32   & $\sim$41-141  & 141   \\
$f_{0}(1500)$ & 1.505$\pm$0.006  & 1.52   & 38$\pm$3      & 38$^*$  \\
$f_{0}(1710)$ & 1.720$\pm$0.006  & 1.70   & $\sim$ 0-6    & 5  \\
$f_{0}(2020)$ & 1.992$\pm$0.016  & 1.88   &  ---          & 0.0  \\
$f_{0}(2100)$ & 2.103$\pm$0.008  & 2.04   &  ---          & 1.2  \\
$f_{0}(2200)$ & 2.189$\pm$0.013  & 2.19   & ---           & 2.5  \\
$f_{0}(2330)$ & 2.29-2.35        & 2.33   &  ---          & 2.8  \\
\hline
\end{tabular} \label{ta1}}
\end{table}
\vspace{-1.1cm}
\section{Conclusions} In this work we obtain a spectrum of high spin mesons, scalar and pseudoscalars in the light-flavor sector,
in agreement to experimental data available using a Dynamical AdS/QCD model. In addition we calculate the decay amplitude of scalar mesons into
two pions. We introduce a mixing angle for $f_0(980)$ of $\pm 20^o$, that corresponds to a composite nature by mixing, e.g., $s\bar{s}$ with light
non-strange quarks\cite{Bediaga}. An absolute value of the mixing angle between $\sim 12^{\circ}$ to $28^{\circ}$ fits $\Gamma^{\pi\pi}$ within
the experimental range. Currently we are including the strange meson sector\cite{AdSQCDStrange} in the Dynamical model. As a future challenge we
also want to introduce finite temperature and calculate the meson spectrum, as done for glueballs within the soft- and hard-wall models
\cite{Miranda}. Finally it could be compared to a large N analysis at finite temperature using lattice simulations recently delivered by
Panero\cite{Panero}.
\section*{Acknowledgments}
We acknowledge partial support from CAPES, FAPESP and CNPq.

\end{document}